\documentclass[useAMS,usenatbib,usegraphicx]{mn2e}
\newcommand\Msun{M$_\odot$}

\newcommand\Ha{H\,{$\alpha$} }
\newcommand\Hb{H\,{$\beta$} }
\newcommand\OIII{[O\,{\sevensize III}] }
\newcommand\OII{[O\,{\sevensize II}] }
\title[\OII emission survey in a cluster XMMXCS~J2215.9--1738]{High star
  formation activity in the central region of a distant cluster at
  {\boldmath$z$}=1.46} 
\author[M. Hayashi et al.]{%
Masao Hayashi,$^{1}$\thanks{E-mail: hayashi@astron.s.u-tokyo.ac.jp}
Tadayuki Kodama,$^{2}$
Yusei Koyama,$^{1}$
Ichi Tanaka,$^{3}$
\newauthor
Kazuhiro Shimasaku,$^{1,4}$
and Sadanori Okamura$^{1,4}$\\
$^{1}$Department of Astronomy, Graduate School of Science, University of Tokyo, Tokyo 113-0033, Japan\\
$^{2}$Optical and Infrared Astronomy Division, National Astronomical Observatory, Mitaka, Tokyo 181-8588, Japan\\
$^{3}$Subaru Telescope, National Astronomical Observatory of Japan, 650 North A'ohoku Place, Hilo, HI 96720, USA\\
$^{4}$Research Center for the Early Universe, Graduate School of Science, University of Tokyo, Tokyo 113-0033, Japan
}

\begin{document}

\date{Accepted 2009 November 10.  Received 2009 September 25; in
  original form 2009 August 2\\} 

\pagerange{\pageref{firstpage}--\pageref{lastpage}} \pubyear{2009}

\maketitle

\label{firstpage}
\begin{abstract}
We present an unbiased deep \OII emission survey of a cluster
XMMXCS~J2215.9-1738 at $z=1.46$, 
the most distant cluster to date with a detection of extended X-ray emission.
With wide-field optical and near-infrared cameras (Suprime-Cam and MOIRCS,
respectively) on Subaru telescope, we performed deep imaging with a narrow-band
filter $NB912$ ($\lambda_c$ = 9139\AA, $\Delta\lambda$=134\AA) as well as
broad-band filters ($B$, $z'$, $J$ and $K_s$).
From the photometric catalogues, we have identified 44 \OII emitters in the
cluster central region of $6'\times6'$ down to a dust-free star
formation rate of 2.6 \Msun/yr (3$\sigma$).  
Interestingly, it is found that there are many \OII emitters even in the
central high density region.
In fact, the fraction of \OII emitters to the cluster members as well as
their star formation rates and equivalent widths stay almost constant
with decreasing cluster-centric distance up to the cluster core. 
Unlike clusters at lower redshifts ($z\la1$) where star formation
activity is mostly quenched in their central regions, this higher
redshift 2215 cluster shows its high star formation activity even at its centre,
suggesting that we are beginning to enter the formation epoch of some galaxies
in the cluster core eventually.
Moreover, we find a deficit of galaxies on the red sequence
at magnitudes fainter
than $\sim$$M^*+0.5$ on the colour-magnitude diagram.
This break magnitude is brighter than that of lower redshift clusters,
and it is likely that we are seeing the formation phase of more
massive red galaxies in the cluster core at $z\sim1$. 
These results may indicate inside-out and down-sizing propagation of star
formation activity in the course of cluster evolution.
\end{abstract}


%%%%%%%%%%%%%%%%%%%%%%%%%%%%%%%%%%%%
\section{INTRODUCTION}
\label{sec;intro}
Galaxy formation and evolution is strongly dependent on environment and
on galaxy mass (e.g., \citealt{dre97,tan05,coo06,tas09}).
Galaxies in high density regions are systematically older than those in
lower density regions, and massive galaxies are on average older than
lower mass galaxies in stellar ages.
It seems that massive galaxies in the high density regions form first
and galaxy formation activity is propagated to lower density regions
and to lower-mass galaxies with time.

Such dependence on environment and mass
can be understood by intrinsic effects and extrinsic effects.
Intrinsic effects are those determined by the initial condition
of galaxy formation and extrinsic effects are those in effect
during the evolution of galaxies after formation.

The environmental dependence can be partly understood by the intrinsic
effect called ``galaxy formation bias'' (e.g., \citealt{cen93})
where high density regions should have started off from the largest
initial density fluctuations that collapse first in the universe,
and galaxy formation takes place earliest in such regions and
subsequent evolution is prompted.
In lower density regions, however, galaxy
formation is delayed and time scales of star formation and mass assembly
are longer.  Likewise, the mass dependence of galaxy formation called
``down-sizing'' (e.g., \citealt{cow96,kod04a}) may be partly
understood by the scaled-down version of the bias on galactic scale,
where more massive galaxies today correspond to higher initial density
fluctuations on galactic scale and their formation such as star
formation and assembly of building blocks take place earlier than 
less massive galaxies.

Galaxies are also subject to external effects from their surrounding environments,
such as galaxy-galaxy interactions/mergers and ram-pressure stripping
in dense environments (e.g., \citealt{aba99,qui00} ).
Such interactions may enhance and/or quench the star formation activity in galaxies
preferentially in high density regions, which would therefore result in conspicuous
dependence of galaxy properties on environments. 

However, the relative importance of the intrinsic effects and the extrinsic effects
is almost totally unknown yet.
One of the most effective methods to verify the existence of the intrinsic effects
is to go back in time and directly see the galaxies in the distant universe.
By doing so, we may eventually reach the epoch when galaxies are forming rapidly
in the biased cluster core, while galaxies are not yet formed or only slowly forming
in lower density regions.  Also, as we go back in time, host galaxies of star
formation would be shifted to higher mass systems and we may eventually see
the active star formation in action in massive galaxies in the cluster core.

In the low redshift universe, star formation
activity is a monotonically decreasing function of local density. 
However, we have not yet known the dependence of star formation activity
on environments at high redshifts in detail.
According to recent studies, it seems that star formation is in fact
biased at high density region at high redshifts. 
\citet{elb07} reported for the first time in the GOODS North/South surveys
that at $z\sim1$ the galaxies at denser environment tend to have
higher star formation rates (SFRs), in contrast to the local Universe.
\citet{coo08} and \citet{ide09} also showed similar trends for \OII
emitters at $z\sim1$ in DEEP2 and COSMOS surveys, respectively.
At a slightly lower redshift, $z=0.81$, based on the mid-infrared observation
of the RX~J1716.4+6708 cluster, \citet{koy08} reported that
the star formation activity is probably enhanced in the medium density
region, such as cluster outskirts or galaxy groups, rather than in the
highest density region.
\citet{pog08} also suggested that SFRs of galaxies may have a
peak at intermediate densities at $z$=0.4--0.8 based on the EDisCS survey. 
These observational findings may imply that the environment that hosts active
star formation is shifted towards denser regions at higher redshifts.

On the other hand, it is well-known as the ``Butcher-Oemler (B-O) effect''
that more distant clusters show higher fraction of blue galaxies up to
$z\sim0.5$, suggesting the enhancement of star formation activity in
higher-$z$ clusters (e.g., \citealt{bo78,bo84,mar01}). However, the
fraction of blue galaxies decreases with cluster centric radius, and red
galaxies still dominate in the core regions at $z\sim0.5$
(\citealt{kod01,ell01}). Such studies of the B-O effect in clusters
have been extended up to $z\sim1$.  \citet{pos01} found that the fraction
of active galaxies in the central regions of clusters is higher at
$z\sim$0.7--0.9 compared to $z\sim$0.2--0.5. However, \citet{nak05}
suggested that the fraction of galaxies with strong \OII emission in the
cluster cores at $z<1.0$ is not significantly dependent on redshift.
It thus seems that we have not yet reached a bursting phase of massive
galaxy formation in cluster cores even at $z\sim1$, although the global
activity of star formation within clusters is already enhanced.

\begin{figure}
\begin{center}
\includegraphics[width=70mm]{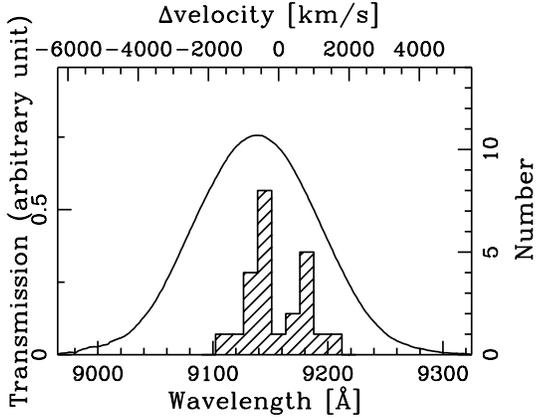}
\end{center}
 \caption{A transmission curve of the $NB912$ filter
   ($\lambda_c$=9139\AA, $\Delta\lambda$=134\AA). Upper x-axis shows
   peculiar velocity with respect to the velocity centre of the
   cluster at $z=1.46$. The histograms show the velocity distribution
   of the spectroscopically confirmed member galaxies of the 2215
   cluster (\citealt{hil09}).
}
 \label{fig;transmission}
\end{figure}

In this paper, we present an \OII emitter survey of the
XMMXCS~J2215.9-1738 cluster (hereafter 2215 cluster)
with a narrow-band filter $NB912$ on Suprime-Cam (Figure \ref{fig;transmission}), 
and discuss spatial distribution of star formation activity
(see also \citealt{koy09}, which presents our \Ha
emitter survey of the RX~J1716.4+6708 cluster at $z=0.81$). 
The 2215 cluster is the most distant cluster to date at $z$=1.46
with a detection of extended X-ray emission (\citealt{sta06}).
\citet{hil07,hil09} have confirmed dozens of the cluster members by
spectroscopy, and found that velocity dispersion of the member galaxies is
$\sigma=580\pm140$ [km s$^{-1}$]. With the $NB912$ filter, we can
survey \OII emissions from the galaxies with the line-of-sight
velocities between $-$2794 $<$ $\Delta V_{\rm los}$ [km s$^{-1}$] $<$
1598 with respect to the velocity centre of the cluster.
Our $NB912$ filter thus perfectly matches this cluster,
and should be able to detect all the \OII emission lines from the
cluster members to a certain flux limit without introducing any bias
(Figure \ref{fig;transmission}).  
Therefore, our survey is unique, and the 2215 cluster is an ideal target
for us to investigate the environmental dependence of star formation activity 
over a wide range in environment at this high redshift.

The X-ray luminosity and inter-cluster medium temperature for the 2215
cluster are 
$L_X=4.4^{+0.8}_{-0.6}\times10^{44}$[erg\ s$^{-1}$], and 
$kT=7.4^{+2.6}_{-1.8}$[keV], respectively (\citealt{sta06}).
The luminosity is fainter than what is expected from the temperature
compared to the local $L_X-T$ relation.  \citet{hil07} thus point out
that this cluster may experience a merger within the last few Gyr. 
It is also found that colour-magnitude diagram in this cluster shows
red sequence (\citealt{sta06,hil09}). \citet{hil09} investigated 
the morphologies of bright member galaxies, and found that
about 60\% of members are E or S0 galaxies.
Even at $z=1.46$, cluster core is already dominated by early-type galaxies,
as far as morphology is concerned.

The structure of this paper is as follows.
The observations and data reduction are described in \S~\ref{sec;obs}.
In \S~\ref{sec;selection}, we describe how we select the \OII emitters
associated to the cluster at $z=1.46$ from the photometric catalogues.
We show the spatial distribution, SFRs, and equivalent widths of the
\OII emitters, and investigate the star forming activity in
the 2215 cluster in \S \ref{sec;results}.  We also discuss the
deficit of faint red galaxies and its evolution.
A summary is given in \S \ref{sec;summary}.
Throughout this paper, magnitudes are in the AB system, 
and we adopt cosmological parameters of $h=0.7$, 
$\Omega_{m0}=0.3$ and $\Omega_{\Lambda 0}=0.7$. 
Vega magnitudes in $J$ and $K_s$, if preferred, can be obtained from
our AB magnitudes using the relations: $J$(Vega)=$J$(AB)$-$0.92 and
$K_s$(Vega)=$K_s$(AB)$-$1.80.

%% figure 1
%%%%%%%%%%%%%%%%%%%%%%%%%%%%%%%%%%%%%%%%%%%%%%%%%%%%%%%%%%%%%%%%%
%%%%%%%%%%%%%%%%%%%%%%%%%%%%%%%%%%%%%%%%%%%%%%%%%%%%%%%%%%%%%%%%%

\section{OBSERVATIONS AND DATA REDUCTION}
\label{sec;obs}
We have obtained optical ($B$, $z'$, and $NB912$) and near-infrared
($J$ and $K_s$) images of the 2215 cluster for selection of \OII
emitters and member galaxies in the cluster. 
The observed data are summarized in Table \ref{table;obs_data}.

\subsection{Optical data}
The optical imaging was conducted using Subaru/Suprime-Cam with
fully-depleted back-illuminated CCDs (\citealt{miy02}) on 2008 July 30
and 31. Suprime-Cam has a wide field-of-view (FoV) of 34$\times$27
arcmin$^2$, and the quantum efficiency of CCDs in red wavelengths, such
as $z'$-band, is considerably improved by replacement with
fully-depleted-type CCDs in the S08A semester (\citealt{kam08}). 
We have obtained two broad-band data of $B$ and $z'$, and a
narrow-band data of $NB912$ (Figure \ref{fig;transmission};
$\lambda_c$=9139\AA, $\Delta\lambda$=134\AA).  
The individual exposure times of a frame in $B$, $z'$ and $NB912$ were
20, 5, and 20 minutes, and the total integration times were
160, 90, and 360 minutes, respectively.
The weather was fine during the two nights of our observing run, and
the sky conditions were photometric all the time, except for the
first half of the night on July 30.
The seeing sizes were 0.6--0.9 arcsec in $B$, 0.8--1.0 arcsec in $z'$, and a
wide range of 0.6--2.0 arcsec in $NB912$. It is noted that the $NB912$
data with bad seeing were taken under marginal condition on July 30.

\begin{figure}
 \vspace{0.5cm}
 \includegraphics[width=75mm]{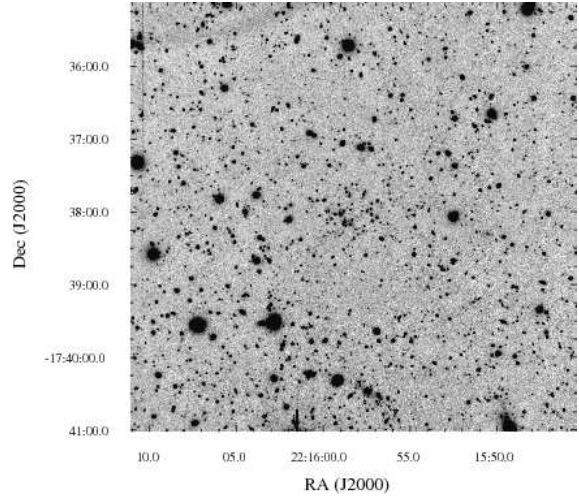}
 \caption{The co-added narrow-band ($NB912$) image of XMMXCS~J2215.9-1738
   cluster taken with Suprime-Cam. North is up, and east is left.
   The field of view is limited to a central $\sim6\times6$ arcmin$^2$
   region, corresponding to a $\sim3\times3$ Mpc$^2$ area, which
   overlaps with our NIR imaging with MOIRCS.
   In this paper, we focus only on this relatively inner region of the
   cluster.
}
 \label{fig;nb912}
\end{figure}

The data reduction is conducted with a data reduction package for
Suprime-Cam (SDFRED ver.1.4: \citealt{yag02}; \citealt{ouc04}). However,
since the currently distributed package is not applicable to the data
obtained after the upgrade of CCDs, we have modified the software
accordingly. The data in each passband are reduced following a standard
procedure: We first rejected some frames with bad seeing or low
signal-to-noise ratio due to high sky background.  A self-flat image
is then created from all the object frames after bias subtraction, and
we use it for flat fielding. A sky is subtracted and a instrumental
distortion is then corrected for on each object frame.  Point spread
functions (PSFs) in all images are matched to 1.09 arcsec, which is
the worst seeing size of the $NB912$ data.  
Finally, after masking satellite trails and bad quality regions, the flux
calibrated images are mosaiced and co-added to make a final image
(Figure \ref{fig;nb912}).  The net integration times in $B$,
$z'$, and $NB912$ are 140, 80, and 260 minutes, and the seeing size in
the final images is 1.09 arcsec.  The zero-point magnitudes are determined
using spectrophotometric standard stars (\citealt{oke90}) for $z'$ and
$NB912$, and a \citet{lan92} photometric standard star for $B$, respectively.
The 3$\sigma$ limiting magnitudes are 27.6, 25.8, and 25.8 in
$B$, $z'$ and $NB912$, respectively.

In this FoV, there are several bright stars that are saturated
especially in $B$. These saturated regions are masked, and are not used in
the following analyses.

%% figure 2
%%%%%%%%%%%%%%%%%%%%%%%%%%%%%%%%%%%%%%%%%%%%%%%%%%%%%%%%%%%%%%%%%
%%%%%%%%%%%%%%%%%%%%%%%%%%%%%%%%%%%%%%%%%%%%%%%%%%%%%%%%%%%%%%%%%

\subsection{Near-infrared data}
\label{sec:nirdata}
The near-infrared (NIR) imaging was carried out with Subaru/MOIRCS
(\citealt{ich06,suz08}) on 2008 June 29 and 30, and we have obtained
images of a central 6$\times$6 arcmin$^2$ area of the 2215 cluster
in two broad-bands, $J$ and $K_s$.
MOIRCS has a FoV of 7$\times$4 arcmin$^2$, which consists of
two chips with a FoV of 4$\times$4 arcmin$^2$ each.
Unfortunately, at the time of our observation, chip 1 was of an
engineering grade, and had a lower quantum efficiency than chip 2 of a
scientific grade. 
We therefore observed at each position with two position angles
of 0 and 180 degrees.  Therefore, four pointings were required
in total to cover the entire $6'\times6'$ region with the chip 2.
The exposure times of a frame in $J$ and $K_s$ are 120 and 40 seconds,
respectively, and three frames were co-added at each dither point in $K_s$.
The total integration times in $J$ and $K_s$ have a range of 32.5--67.5
and 18--30 minutes, respectively, depending on the pointing. The sky
conditions were photometric during the two nights of our observing run.
The seeing sizes were 0.4--0.7 arcsec in $J$ and 0.4--0.8 arcsec in $K_s$.

The data reduction is conducted with a data reduction package for
MOIRCS (MCSRED ver.20081023: Tanaka et al. in preparation). The data
are reduced following a standard procedure: A flat-fielding is done
using a self-flat image created from the object frames. A sky is
subtracted and instrumental distortion is then corrected for on each
image. 
In the MOIRCS frames, especially those taken by chip 2,
strong fringe pattern are sometimes seen, which are removed from
the images.
We masked some regions with low signal-to-noise ratio such as the
edges of the frames, matched the PSF sizes, and mosaiced and co-added
to make the final images.  The seeing sizes in $J$ and
$K_s$ images are degraded to 1.09 arcsec to match the optical images.
The zero-point magnitudes are determined using a faint standard star
in \citet{leg06} that we observed during our run.
The 3$\sigma$ limiting magnitudes are 23.8--24.6, and
23.1--23.6 in $J$ and $K_s$, respectively. 

In what follows, we focus only on the central region (33.8 arcmin$^2$)
of the 2215 cluster (Figure \ref{fig;nb912}) where both NIR and
optical data exist, because the NIR data are crucial to identify
member galaxies of the cluster.

%% table 1
%%%%%%%%%%%%%%%%%%%%%%%%%%%%%%%%%%%%%%%%%%%%%%%%%%%%%%%%%%%%%%%%%
\begin{table}
 \caption{Summary of the observed data. The limiting magnitudes are
 measured with a 2\arcsec diameter $\phi$.  Depth of the NIR data
 depends on the pointing due to different integration times.
 The seeing sizes in all the co-added images are matched to 1.09 arcsec.
 In this paper, we will concentrate only on the central 6$\times$6
 arcmin$^2$ region where the NIR data co-exist.
  }
\begin{center}
\begin{tabular}{ccccc}
\hline\hline
filter & effective FoV & net integration & limiting mag. \\
       & (arcmin$^2$)  & (minutes)    & (3$\sigma$)   \\
\hline
$B$    & 32$\times$23  & 140 & 27.6 \\
$z'$   & 32$\times$23  & 80  & 25.8 \\
$NB912$& 32$\times$23 & 260 & 25.8 \\
$J$    & 6$\times$6 & 32.5 -- 67.5 & 23.8 -- 24.6 \\
$K_s$  & 6$\times$6 & 18 -- 30 & 23.1 -- 23.6 \\
\hline\hline
\label{table;obs_data}
\end{tabular}
\end{center}
\end{table}
%%%%%%%%%%%%%%%%%%%%%%%%%%%%%%%%%%%%%%%%%%%%%%%%%%%%%%%%%%%%%%%%%

%% figure 3a, 3b
%%%%%%%%%%%%%%%%%%%%%%%%%%%%%%%%%%%%%%%%%%%%%%%%%%%%%%%%%%%%%%%%%
\begin{figure*}
 \includegraphics[width=85mm]{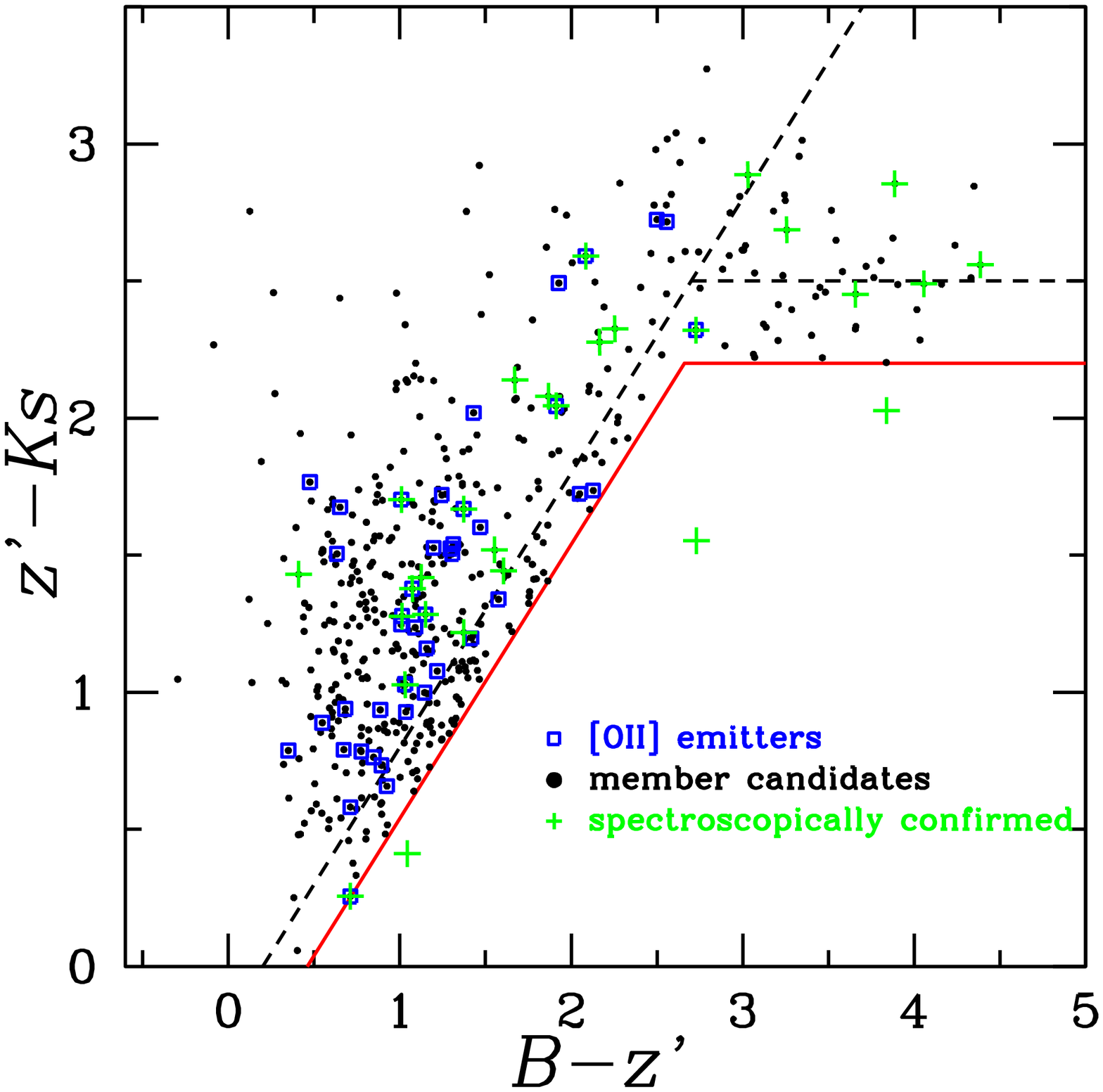}
 \includegraphics[width=85mm]{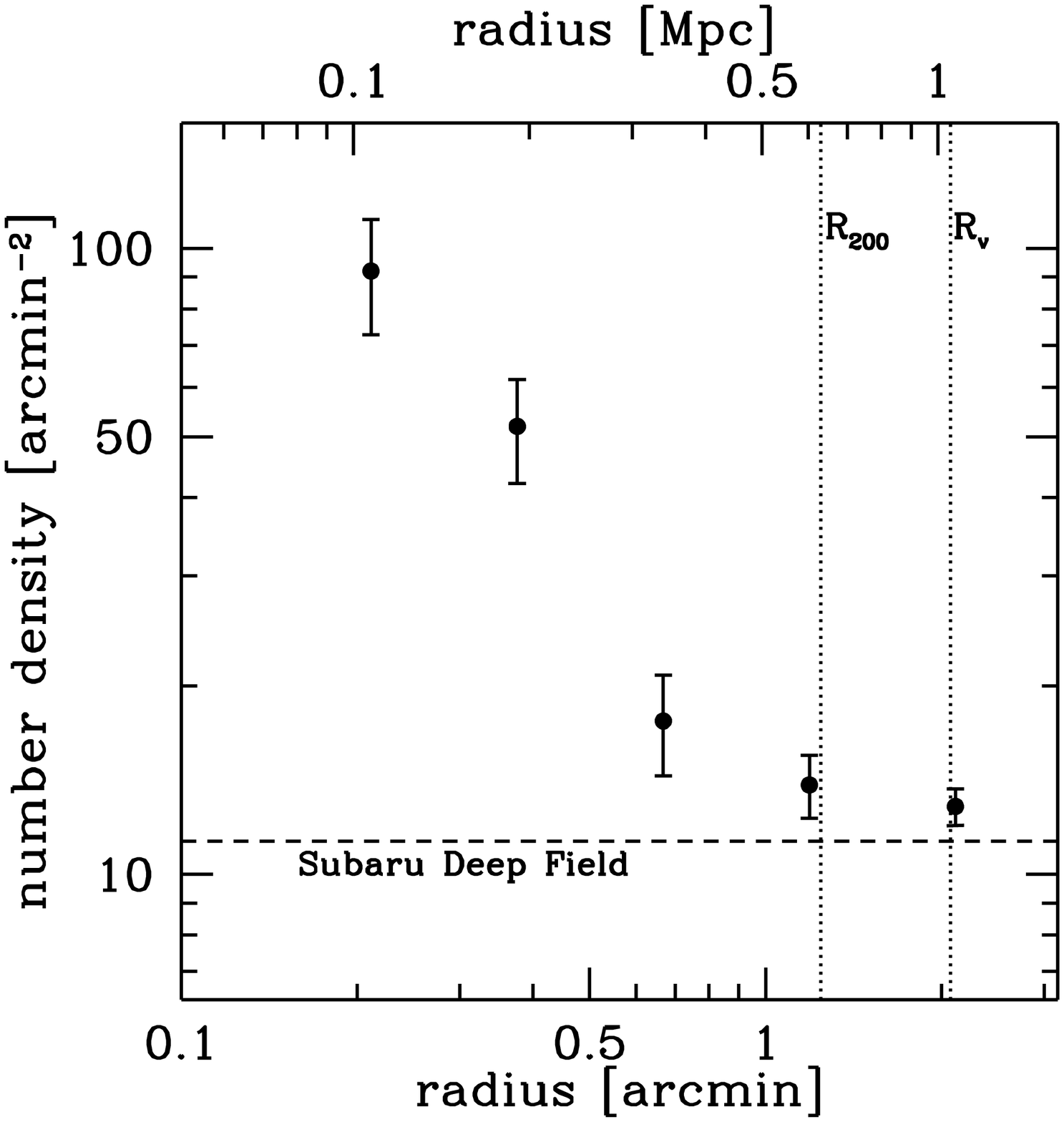}
 \caption{(a) Left panel: The colour-colour diagram of $B-z'$
   vs. $z'-K_s$.  The red solid lines show our selection criteria for
   cluster member candidates, while the black broken lines indicate
   the original BzK selection criteria defined by \citet{dad04}.
   Blue open squares show 44 \OII emitters in the 2215 cluster, black
   dots are cluster member candidates, and green crosses indicate 
   spectroscopically confirmed cluster members in \citet{hil07,hil09}.
   (b) Right panel: The number density of cluster member candidates as
   a function of radius from the cluster centre.  Error bars show Poisson
   errors.  Vertical lines show the radii of $R_{200}=0.63$Mpc and 
   $R_{v}=1.05$Mpc, respectively (\citealt{hil07}). Horizontal line
   indicates the number density of Subaru Deep Field galaxies selected
   with the same criteria.
}
 \label{fig;bzk}
\end{figure*}
%%%%%%%%%%%%%%%%%%%%%%%%%%%%%%%%%%%%%%%%%%%%%%%%%%%%%%%%%%%%%%%%%

\section{SELECTION OF CLUSTER MEMBERS AND [O\,{\sevensize\bf II}] EMITTERS}
\label{sec;selection}
\subsection{Catalogues}
\label{sec;selection_nbcatalogue}
\subsubsection{$NB912$-detected catalogue}
We use SExtractor (ver. 2.5.0: \citealt{ber96}) for source detection and
photometry.  Objects are detected in $NB912$, and their photometries
in all images are carried out at the same positions as on the $NB912$
image using the double-image mode of SExtractor. All images are
geometrically matched to the $NB912$ data. 
We then select 1575 objects brighter than 5$\sigma$ limiting magnitude
in $NB912$, where objects in masked regions are excluded.  Color indices
are derived within a 2\arcsec-diameter aperture, and MAG\_AUTO
magnitudes are used as total magnitudes.

Stars and galaxies are separated based on their $B-z'$ and $z'-K_s$
colours according to the same procedure as in \citet{hay07} (see also
\citealt{dad04}, \citealt{kon06}).  
Among 1575 objects, 158 are classified as stars, and the remaining
1417 are classified as galaxies. Using the stellar sample, we check the zero points of 
magnitudes in all the optical and NIR images by comparing stellar
colours with those of stellar spectrophotometric atlas of 
\citet{gun83}.
No correction is applied for all the bands since stellar colours
are in good agreement with those of the stellar atlas,
except for the $z'$-band in which we shifted the zero-point
by only 0.05 magnitude.

Next, we compare differences between aperture magnitude and total
magnitude in all bands for galaxy sample, and find that aperture
magnitudes are a little fainter systematically than total magnitudes
in only $z'$-band. Therefore, aperture correction of $0.05$ magnitude
is applied for aperture magnitudes in $z'$ so that an aperture
contains the same fraction to total in all bands.
In addition, magnitude is corrected for galactic absorption,
A($B$)=0.10, A($z'$)=0.04, A($NB912$)=0.04, A($J$)=0.02, and
A($K_s$)=0.01, which are estimated using extinction law of
\citet{car89} on the assumption of $R_V$=3.1 and $E(B-V)$=0.025 at the
central coordinates of the 2215 cluster according to \citet{sch98}.

Magnitude errors are estimated from 1$\sigma$ sky noise taking account
of the difference in depth at each object position due to
slightly different exposure times and sensitivities.

\subsubsection{$K_s$-detected catalogue}
We also create a $K_s$-detected catalogue using the same procedure as
we used for the $NB912$-detected catalogue, except that $K_s$-band image
is used instead for source detection. We then select 1198 objects
(115 stars and 1083 galaxies) brighter than $K_s$=23.0, which corresponds
to $\sim$5$\sigma$ limiting magnitude in most of the regions and
$\sim$3$\sigma$ in the shallowest region.

The $K_s$-detected catalogue is used only in \S \ref{sec;cmd} for discussion
based on the $z'-K_s$ vs. $K_s$ colour-magnitude relation of the cluster members.
For the rest of the paper, the $NB912$-detected catalogue is used unless otherwise
mentioned.

\subsection{Cluster member candidates}
\label{sec;clmember}
It would be ideal of course to use spectroscopic redshifts to determine membership
of individual galaxies.  The lack of such complete spectroscopic data, however,
forces us to estimate membership based on our existing photometries.
In this paper, we use a $B-z'$ vs. $z'-K_s$ colour-colour diagram,
in analogy to the BzK selection of galaxies \citep{dad04}.
We note that our scheme is very simple and easily reproducible,
unlike commonly used photometric redshifts based on spectral fitting to some model
or empirical templates.
We apply the following colour criteria, 
\begin{equation}
(z'-K_s)>(B-z')-0.46\ \cup\ (z'-K_s)>2.2,
\label{eq;member}
\end{equation}
in order to remove the foreground and background galaxies
(red solid lines in Figure \ref{fig;bzk}(a)).
We modify the original BzK criteria (broken lines in Figure
\ref{fig;bzk}(a); \citealt{dad04})
that has been designed to select both star-forming and passively
evolving galaxies at $1.4 \la z \la 2.5$.
Since the redshift under concern in this paper, namely the cluster
redshift of 1.46, is close to the lowest edge of the applicable
redshift range of the BzK selection, we loosen the original boundary
of the BzK selection as indicated by solid lines in
Figure~\ref{fig;bzk}(a) to increase the completeness of galaxies at
$z=1.46$. Consequently, it is also obvious that our criteria will
select a large number of background galaxies up to $z\sim2.5$ which
are contaminants for our use. 
However, for the narrow-band emitters in $NB912$, we just need to separate
out the \OII line emitters from other line emitters such as H$\alpha$ and \OIII
in the foreground and background at some specific redshifts.
These unwanted emitters are well separated on this diagram.
For the non-emitters, we will apply a statistical correction to the above
defined cluster member candidates, using a blank field data in the
Subaru Deep Field (see below).

We use the $NB912$-detected catalogue to select cluster member
candidates.  If an object is not detected at more than 2$\sigma$ above
the noise level in each band, its magnitude is converted to the 2$\sigma$
limiting magnitude, so that the upper/lower limit of colour can be derived.
For membership, we only pick out the galaxies that strictly meet our colour
criteria (the principle of our selection is similar to that used in \citealt{hay07}).
Since the colour selection based on the BzK diagram for objects
without secure $K_s$-band detection is uncertain, we exclude those
fainter than the 2$\sigma$ limiting magnitude in $K_s$ in the
membership determination.
We select 486 cluster member candidates from the $NB912$-detected
catalogue, among which 365 galaxies meet the original BzK criteria.
The number density is 14.4 per arcmin$^2$. 

In our colour-selected member candidates, there must be a significant
contamination of foreground/background galaxies that are not associated
to the 2215 cluster.
We estimate a contribution of such field galaxies by applying the same
criteria and limiting magnitudes to a part of the Subaru Deep Field (SDF)
of a 453 arcmin$^2$ area which has deep $K$ imaging data as well (\citealt{mot08}).
The number density of the colour-selected galaxies is 11.3 per arcmin$^2$,
which is in good agreement with the number density in the outer region
of the cluster (Figure \ref{fig;bzk}(b)).
The excess over the field density can be regarded as a cluster component.
This implies that our estimations of the number density of field
contamination, thus that of cluster members, are reasonable.

\citet{hil09} reported 64 cluster members within the radius
of 1.5 arcmin in the 2215 cluster. Among them, 24 galaxies are
identified by spectroscopy, and the others are selected by
photometric redshifts. The spectroscopically confirmed members
are also plotted in Figure \ref{fig;bzk}(a).  We find that $\sim$90\%
of members in \citet{hil09} are contained in our member candidate
sample, which assures our member selection.  Although three
confirmed members are missed out in our selection, such low-level
incompleteness would not significantly affect our statistical
discussions in this paper.
We then use our sample to estimate the number of cluster members after
subtracting the field contribution, and find that our estimation of
$\sim79\pm15$ members is compatible with the number in \citet{hil09}.
This fact also supports the validity of our criteria for selection of
cluster members.

\subsection{[O\,{\sevensize\bf II}] emitters}
\label{sec;oiiemitters}

%% figure 4
%%%%%%%%%%%%%%%%%%%%%%%%%%%%%%%%%%%%%%%%%%%%%%%%%%%%%%%%%%%%%%%%%
\begin{figure}
 \begin{center}
 \includegraphics[width=70mm]{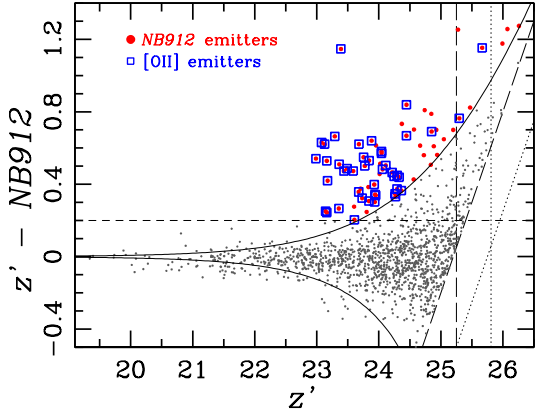}
 \end{center}
 \caption{A colour-magnitude diagram of $z'$ vs. $z'-NB912$. Solid lines
 show the 3$\sigma$ excesses of $NB912$ over $z'$. Long-dashed and dotted lines
 show 3$\sigma$ and 5$\sigma$ limiting magnitudes,
 respectively. Short-dashed line shows $z'-NB912=0.2$. Gray dots show
 galaxies brighter than 5$\sigma$ in $NB912$. Red filled circles are
 69 $NB912$ emitters, and blue open squares are 44 \OII emitters in
 the 2215 cluster.}
 \label{fig;zznb}
\end{figure}
%%%%%%%%%%%%%%%%%%%%%%%%%%%%%%%%%%%%%%%%%%%%%%%%%%%%%%%%%%%%%%%%%

We select galaxies with flux excess in $NB912$ compared to $z'$
to search for \OII emitters in the 2215 cluster.
The \OII line ($\lambda_{\rm rest}$=3727\AA) emitted by galaxies
at $z=1.46$ is redshifted just into the $NB912$ filter, while
the $z'$-band samples averaged continuum flux underneath the line.
This survey based on $z'-NB921$ colours is therefore best suited to
this 2215 cluster and is able to search for \OII emissions from
almost all the member galaxies of the cluster to a certain flux limit.
The galaxies with colour excesses in $z'-NB912$ larger than 3$\sigma$ 
and $z'-NB912>0.2$ are identified as the $NB912$ emitters.
(Figure \ref{fig;zznb}).
The 3$\sigma$ colour of $z'-NB912$ is derived as follows; 
\begin{equation}
3\sigma_{z'-NB912}=-2.5\log[1\pm\sqrt[]{\mathstrut f^2_{3\sigma,z'}+f^2_{3\sigma,NB912}}/f_{z'}],
\end{equation}
where $f_{3\sigma}$ is 3$\sigma$ sky noise flux. 
The latter criterion is arbitrarily determined from the distribution
of galaxy colours.  However, $z'-NB912$ colours of spectral templates of
local galaxies (Coleman, Wu \& Weedman 1980) redshifted over the range of $z$=0.0--3.0
indicate that no galaxy would satisfy $z'-NB912>0.2$ without having any
emission line in $NB912$, suggesting we have no contamination from ordinary galaxies. 
The colour criteria correspond to \OII flux $\gid$ 1.4$\times$10$^{-17}$ erg
s$^{-1}$ cm$^{-2}$ and observed equivalent width $\gid$ 35\AA,
respectively. This means that our survey is sensitive to the \OII emitters
with a dust-free SFR larger than 2.6 \Msun yr$^{-1}$ according to the
the \OII--SFR calibration in \citet{ken98}. 

In order to accurately measure the flux excess in $NB912$ to $z'$,
we may want to take into account the difference in continuum flux density, if any,
between $z'$ and $NB912$ bands due to the slope of continuum
spectrum.  However, if we compare $z'-NB912$ vs. $z'-K_s$, we found
that there is no correlation between the two colours, and that $z'-NB912$
colours are distributed around $z'-NB912=0$ with a certain dispersion.
We therefore conclude that the difference in continuum between $z'$ and
$NB912$ is negligible.
This is also supported by the fact that the difference in effective
wavelengths between the two bands is small;
$\Delta\lambda\sim100$\AA.  Therefore no correction for the colour term
is made.  After all, we select 69 $NB912$ emitters from 1417 galaxies
in the $NB912$-detected sample. Out of the 69 emitters, 55 are
detected in $K_s$ at more than 2$\sigma$ levels.

\begin{figure}
 \begin{center}
 \includegraphics[width=70mm]{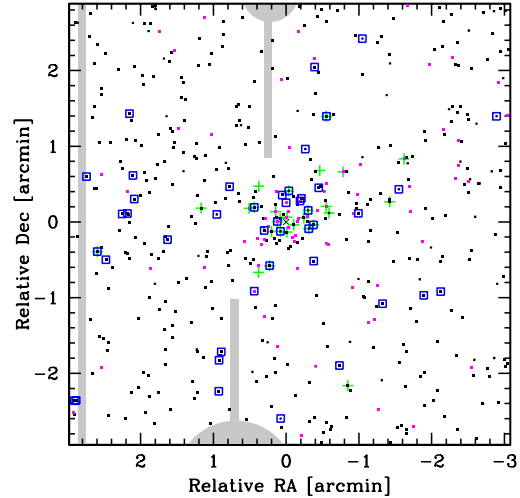}
 \end{center}
 \caption{The celestial distribution of 44 \OII emitters and cluster
   member candidates in the 2215 cluster. North is up, and east is
   to the left. Symbols are the same as those in Figure~\ref{fig;bzk}(a),
   except that magenta dots show red member candidates that satisfy
   $z'-K_s>-0.088K_s+4.26$ (lower broken line in
   Figure~\ref{fig;cmd_lf}(a); see section~\ref{sec;cmd}).
   Gray areas indicate masked regions that we exclude from the analyses.
}
 \label{fig;map_member}
\end{figure}

Our $NB912$ emitters could include galaxies with the
following major strong lines such as
\Ha ($\lambda_{\rm rest}$=6463\AA, $z$=0.39), \OIII 
($\lambda_{\rm rest}$=4959 and 5007\AA, $z$=0.82--0.84), \Hb
($\lambda_{\rm rest}$=4861\AA, $z$=0.88), \OII 
($\lambda_{\rm rest}$=3727\AA, $z$=1.46) and Ly$\alpha$ 
($\lambda_{\rm rest}$=1216\AA, $z$=6.51).
We then apply our selection criteria of cluster member candidates
defined in \S\ref{sec;clmember} for the 55 $NB912$ emitters, of which
44 emitters are identified as \OII emitters (Figure
\ref{fig;zznb}). Figure \ref{fig;map_member} shows a spatial
distribution of those 44 \OII emitters.

%% figure 5
%%%%%%%%%%%%%%%%%%%%%%%%%%%%%%%%%%%%%%%%%%%%%%%%%%%%%%%%%%%%%%%%%
%%%%%%%%%%%%%%%%%%%%%%%%%%%%%%%%%%%%%%%%%%%%%%%%%%%%%%%%%%%%%%%%%

The empirical colour tracks between $z=0.0$ and 1.5 on the
$B-z'$ vs. $z'-K_s$ diagram based on the spectral templates
(\citealt{cww80}) assure that most of the galaxies at $z<1.0$ cannot
satisfy the colour selection of equation (\ref{eq;member}). 
Moreover, almost all the $NB912$ emitters are detected in $B$-band at
more than 5$\sigma$ confidence level, which would reject the
possibility that the emission line is Ly$\alpha$ at $z=6.51$.
Therefore, by applying the modified BzK selection, we can securely
decontaminate other lines than \OII at $z$=1.46.

We also use the samples of \Ha ($z=0.4$), \OIII ($z=0.84$) and \OII ($z=1.47$)
emitters identified from the $NB921$ emitters in the SDF
(\citealt{ly07}) in order to verify our selection of \OII emitters.
\citet{ly07} separated \Ha emitters based on $R_c-i'$ vs. $B-R_c$ colours,
and then discriminated between \OIII and \OII emitters based on $i'-z'$ vs. $R_c-i'$
colours, as well as some spectoroscopic data where available.
Although our method uses less colour information, i.e. $B-z'$ and $z'-K_s$ only,
it is found that our criteria can still effectively separate \OII lines
from other possible lines.

We compare the number densities between our emitter samples and the
SDF emitter samples (\citealt{ly07}) under the same selection criteria
and limiting magnitude.
We assume that the remaining 11 $NB912$ emitters other than \OII in our
sample are most likely to be strong \Ha or \OIII lines, and compare
its number density, 0.33$\pm$0.10 per arcmin$^2$, with that of a
combined sample of \Ha and \OIII lines in the SDF, 0.30$\pm$0.02 per
arcmin$^2$. Since these emitters do not belong to the 2215 cluster at
$z=1.46$, this agreement in the number densities supports the validity
of our line identification.

The number density of \OII emitters in the cluster, 1.3$\pm$0.2 per arcmin$^2$,
is a factor of $\sim$4 larger than that of the SDF sample, 0.34$\pm$0.02 per arcmin$^2$.
Such excess are probably due to the existence of many \OII emitters
associated to the cluster at $z$=1.46.
A cosmic variance due to our relatively small FoV ($6'\times6'$) of this study
may contribute to the difference to some extent.
We examine the number of \OII emitters within randomly allocated $6'\times6'$
FoVs in the SDF, and find that the number of 44 \OII emitters in the cluster
is $\sim$3.6$\sigma$ excess than median field number density of \OII emitters.
This suggests that the difference is intrinsic and unlikely to be caused
only by such cosmic variance.

\section{RESULTS AND DISCUSSIONS}
\label{sec;results}
\subsection{Star formation activity in a cluster core}
\label{sec;sf_activity}

%% figure 6
%%%%%%%%%%%%%%%%%%%%%%%%%%%%%%%%%%%%%%%%%%%%%%%%%%%%%%%%%%%%%%%%%
\begin{figure}
 \begin{center}
 \includegraphics[width=70mm]{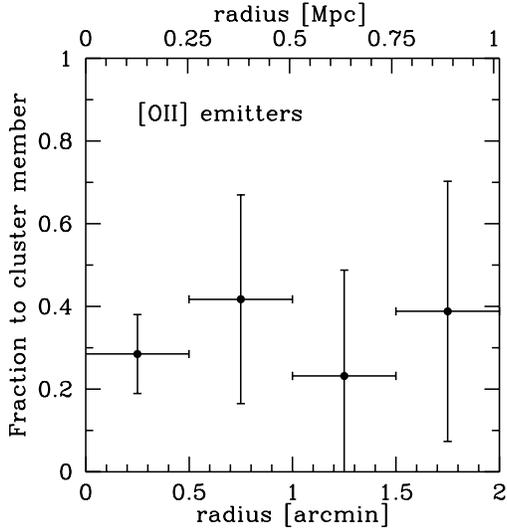}
 \end{center}
 \caption{Fraction of \OII emitters to the cluster members as a function of
 radius from a cluster centre. The contribution of field galaxies is
 statistically subtracted from our colour-selected cluster member candidates.
 The error-bars indicate the statistical errors associated to both \OII
 emitters and the cluster members.
}
 \label{fig;oii_fraction}
\end{figure}
%%%%%%%%%%%%%%%%%%%%%%%%%%%%%%%%%%%%%%%%%%%%%%%%%%%%%%%%%%%%%%%%%

%% figure 7
%%%%%%%%%%%%%%%%%%%%%%%%%%%%%%%%%%%%%%%%%%%%%%%%%%%%%%%%%%%%%%%%%
\begin{figure}
 \begin{center}
 \includegraphics[width=84mm]{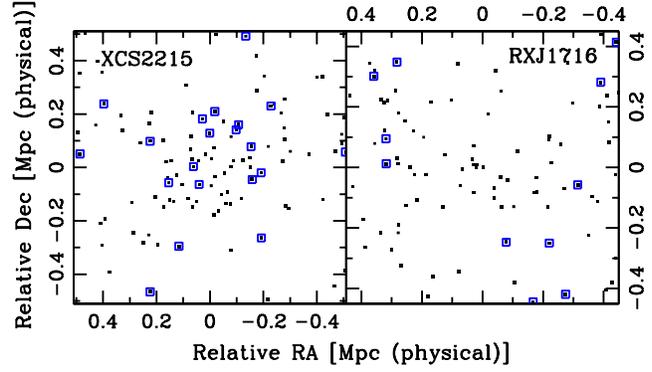}
 \end{center}
 \caption{Close-up views of the central $2'\times2'$ regions for
 XCS2215 cluster at $z=1.46$ (left) and RXJ1716 cluster at $z=0.81$
 (right). Blue open squares show \OII emitters for 2215 cluster and
 \Ha emitters for RXJ1716 cluster, respectively. Dots show cluster
 member candidates selected in \S\ref{sec;clmember} for the 2215
 cluster, and member galaxies selected by photometric redshifts of
 $\Delta z=0.76-0.83$ for RXJ1716 cluster (\citealt{koy07}),
 respectively. Both panels show the areas of similar physical scales
 (0.51Mpc/arcmin at $z=1.46$ and 0.45Mpc/arcmin at $z=0.81$,
 respectively). 
}
 \label{fig;1716}
\end{figure}
%%%%%%%%%%%%%%%%%%%%%%%%%%%%%%%%%%%%%%%%%%%%%%%%%%%%%%%%%%%%%%%%%

%% figure 8a, 8b, 8c
%%%%%%%%%%%%%%%%%%%%%%%%%%%%%%%%%%%%%%%%%%%%%%%%%%%%%%%%%%%%%%%%%
\begin{figure*}
 \includegraphics[width=55mm]{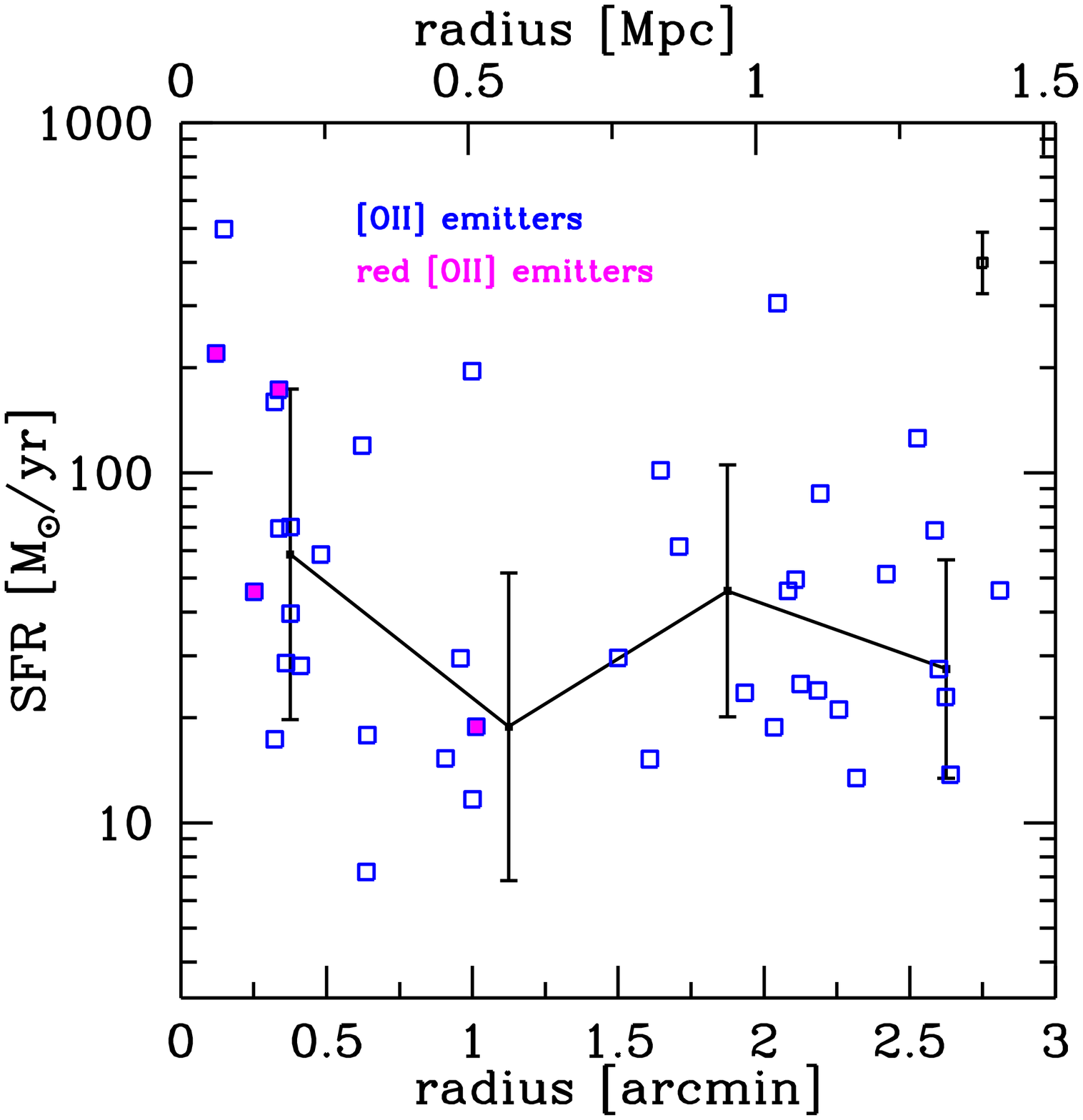}
 \includegraphics[width=55mm]{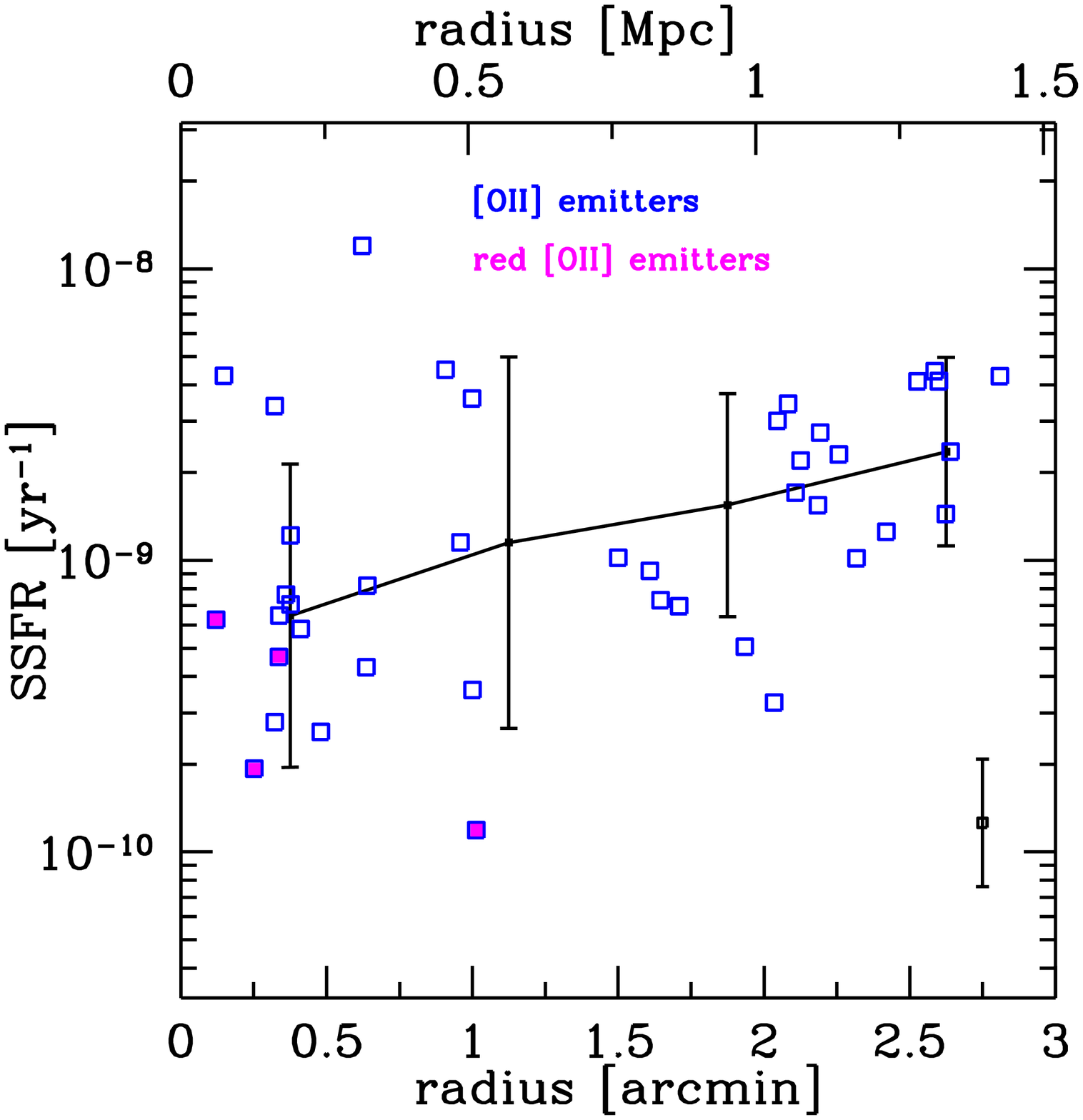}
 \includegraphics[width=55mm]{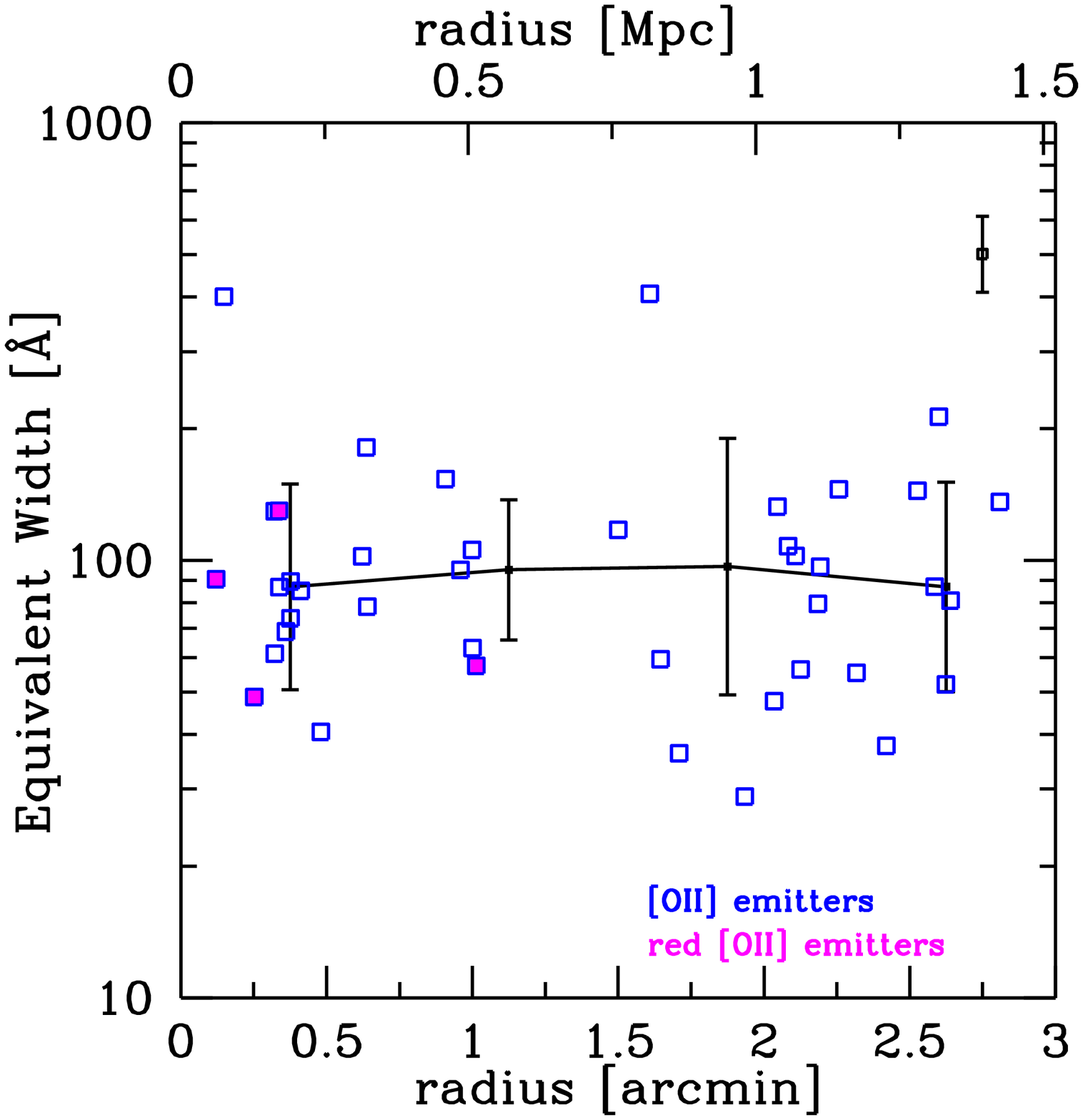}
 \caption{
 SFRs(left), specific SFRs (SSFRs) (middle) and observed
 equivalent widths (right) for \OII emitters as a function
 of distance from the cluster centre.
 SFRs are derived and corrected for dust extinction using
 the relations in \citet{mou06} (see text).
 Stellar masses are estimated using the
 equation of \citet{dad04}. Equivalent widths are calculated with the
 equations (\ref{eq;fline}) and (\ref{eq;fcont}). Blue open squares
 show 44 \OII emitters, and 4 red \OII emitters are marked by magenta
 filled square. Typical errors on each value are shown in the upper or
 the lower right corner on each panel. The errors on SFR and equivalent
 width are derived from magnitude errors in $NB912$ and $z'$, and
 $\sigma(\Delta \log({\rm M_\star}))=0.20$ is applied to the error in
 stellar mass (see text). 
 In each diagram, the solid line connects
 median values in four radius bins, and the errors contain both the
 typical error on each value and the standard deviation in each bin.
}  
 \label{fig;sfr_ew}
\end{figure*}
%%%%%%%%%%%%%%%%%%%%%%%%%%%%%%%%%%%%%%%%%%%%%%%%%%%%%%%%%%%%%%%%%

Figure \ref{fig;map_member} shows that there are many \OII emitters in
the central region of the 2215 cluster.
If we assume that \OII lines are emitted from ionized gas in/around
the star forming regions, it seems that the 2215 cluster is still
actively forming stars even at its core region.
This is not the case in lower-$z$ clusters where star forming activity
is much lower in the central region.
To be more quantitative, we estimate a fraction of \OII emitters to
cluster members as a function of cluster-centric distance
in Figure~\ref{fig;oii_fraction}.
Here, the amount of remaining contamination from foreground/background
galaxies are estimated from the SDF sample, and it is statistically
subtracted from the colour-selected sample of cluster member candidates.
Figure \ref{fig;oii_fraction} suggests that the 2215 cluster 
maintains a high fraction of star forming galaxies in the central
region, $\sim 30$\%, even at the most inner part within 0.25~Mpc
in physical scale (1'=0.51Mpc at $z=1.46$).
This fraction is higher than the \OII fraction in the cluster cores
in \citet{nak05}, which is $\la 20$\% at $z<1.0$.

\citet{lid08} recently reported a distribution of spectroscopic sample of
\OII emitters in a cluster XMMU~J2235.3-2557 at $z=1.39$, a similar
redshift to that of the 2215 cluster.
While there are no galaxies in the very core within 90~kpc in radius
that show star forming activity, \OII lines are detected from more
than half of galaxies near the centre.
This result is similar to ours on the \OII emitters in 2215 cluster
at a similar (slightly higher) redshift of 1.46.
It seems likely that distant X-ray detected clusters at $z\ga1.4$
are actively forming new stars even in the central region within
a few hundreds kpc in radius.

We have recently conducted a narrow-band \Ha emitter survey for
RX~J1716.4+6708 cluster at $z$=0.81 (\citealt{koy09} in preparation).
This survey reveals quite a different situation where no \Ha emitters
are observed in the core region within a radius of 0.25~Mpc (Figure \ref{fig;1716}).
A difference in spatial distribution of the emitters is very impressive.
The \Ha emitters in the 1716 cluster are selected with a combination
of $NB119$ ($\lambda_c$ = 11885\AA, $\Delta\lambda$=141\AA) and $J$
filters on Subaru/MOIRCS.
The \Ha survey reached to the depth of a limiting line flux of
$\sim$4.1$\times$10$^{-17}$ erg s$^{-1}$ cm$^{-2}$, which corresponds
to a dust-free SFR of $\sim1.0$\Msun yr$^{-1}$.
The \Ha survey for the 1716 cluster is, thus, more sensitive to star forming
galaxies with slightly lower SFR than the \OII survey for the 2215
cluster, presented in this paper.

Although the lines are different between the two clusters, \Ha and
\OII lines, considering the fact that \Ha line is much less affected
by dust extinction than \OII line, the intrinsic difference in spatial
distribution of the emitters would be more significant.
Furthermore, \citet{koy09} find that the fraction of star forming
galaxies decreases as one goes to denser region, which is different from
what we see for the 2215 cluster.  Similar trends are also found in
lower redshift clusters at $z$=0.4--0.8 (\citealt{kod04b,pog08}).
\citet{koy08,koy09} have also unveiled the star forming activities
hidden by dust in the 1716 cluster based on the mid-infrared imaging
with AKARI.
These studies conclude that star formation activity has already been
quenched in the cluster core, while it comes to an peak in the medium
density regions away from the cluster core.

These facts may imply that we find galaxy formation bias in the
highest density region at $z\sim1.5$.  Recent studies also support the
biased star formation in a relatively dense environment at $z>1$ 
(\citealt{elb07,coo08,ide09}).
The difference of critical environments in star formation at various
redshifts may suggest that galaxy formation bias plays an important
role in the dependence of galaxy properties on environments.

In such a comparison between different clusters at different redshifts,
we must keep in mind that the size and mass of the clusters can be different.
In fact, the bolometric X-ray luminosity of 1716 cluster is $\sim3$
times larger than that of 2215 cluster (\citealt{ett04,sta06}).
Also, the sub-structures of 1716 cluster is more prominent (\citealt{koy07}).
We cannot therefore conclude whether the difference seen in spatial
distribution is largely due to time evolution or due to different masses or
characteristics.
But it may be the case that we are witnessing the evolution in star forming
activity in the core of high-$z$ clusters from $z\sim1.5$ to $z\sim0.8$.

On the other hand, \citet{fin05} report a diversity in distribution of
\Ha emitters in the cores of clusters at $z$=0.7--0.8.
Two among their three clusters at similar redshifts show a decrease
in \Ha fraction toward the cluster centre, while the other
cluster shows an opposite trend.
This fact may imply that it is not only the star formation bias that
causes the high fraction of \OII emitters in the core of the 2215 cluster.
It is thus crucial to observe more clusters at $z>1.0$, and to evaluate
to what extent the presence of star forming galaxies in the cluster
core is a general property at $z>1$.

The $L_X-T$ relation of the 2215 cluster suggests that it is possible
that it experienced a merger event within the last few Gyr
(\citealt{hil07}). This can be another possible reason for the high
fraction of \OII emitters, since the cluster merging event might cause
an enhancement of star formation activity in galaxies in the cluster core.
Also, since the distribution of \OII emitters is projected on
celestial sphere, it may be possible that some member galaxies in the 
outskirt of the cluster happen to be seen in the direction to the
cluster centre, which may result in apparent high fraction of \OII
emitters at the centre.
However, the number density of galaxies in the outskirt is likely to be
much lower than that of galaxies in the core region (Figure \ref{fig;bzk}(b)),
and such projection effect would be small in the direction to the centre.

It is also possible that an active galactic nucleus (AGN) enhances
\OII line flux.  Recent studies suggest that a fraction of AGNs in
clusters increases with redshifts.  \citet{gal09} have found an
overdensity of AGNs within a radius of 0.25~Mpc in clusters at $z>0.5$,
and that the density of X-ray selected AGNs in clusters at $1.0<z<1.5$
is 0.102 arcmin$^{-2}$, which is $\sim2$ times larger than that of
clusters at $0.5<z<1.0$.  It indicates that it is possible that 3--4
AGNs reside in the central 33.8 arcmin$^{2}$ region of the 2215 cluster. 
\citet{mar09} also show a monotonous increase of AGN fraction in
clusters from $z\sim0.2$ to $z\sim1.0$.
It is worth mentioning that \citet{sta06} report no obvious X-ray point
source in the cluster core, although we cannot completely exclude the
presence of any point source. We thus consider that AGN contribution is small.
\citet{yan06} suggest that \OII emitters in the red sequence tend to be
AGNs, most commonly LINERs, and that \OII line can be
valid as an indicator of star formation activity only for blue
galaxies. We find four red \OII emitters with $z'-K_s$ colours redder
than $-0.088K_s+4.26$ (see \S\ref{sec;cmd} and Figure \ref{fig;cmd_lf}(a)), 
which are located near the centre (Figure \ref{fig;map_member}). 
Even if all these red \OII emitters are AGNs, the enhancement of
galaxy activity in the core would be still valid, since the presence of AGN
is also a sign of activity.  This would then provide us with the clues to
understanding the influence of AGNs on the evolution of massive galaxies
in the cluster core, such as quenching of star formation.
It is essential therefore to obtain NIR spectra of these four \OII
emitters in order to constrain the origin of the \OII lines.

\subsection{SFR, specific SFR and equivalent width}
According to \citet{ly07}, \OII line flux [ergs\ s$^{-1}$ cm$^{-2}$]
and continuum flux density [ergs\ s$^{-1}$ cm$^{-2}$ \AA$^{-1}$] for
44 \OII emitters are calculated from flux densities in $NB912$ and
$z'$ bands ($f_{NB912}$ and $f_{z'}$), respectively, as follows;
\begin{equation}
F({\rm [O\,{\scriptstyle
      II}]})=f_{NB912}\Delta_{NB912}\frac{1-(f_{z'}/f_{NB912})}{1-(\Delta_{NB912}/\Delta_{z'})},
\label{eq;fline}
\end{equation}
\begin{equation}
f_{\lambda,{\rm cont}}=f_{z'}\frac{1-(f_{NB912}/f_{z'})(\Delta_{NB912}/\Delta_{z'})}{1-(\Delta_{NB912}/\Delta_{z'})},
\label{eq;fcont}
\end{equation}
where $\Delta_{NB912}$ and $\Delta_{z'}$ indicate FWHMs of the filters,
and $\Delta_{NB912}=134$\AA\ and $\Delta_{z'}=955$\AA. 

In order to estimate dust-corrected SFR of our \OII emitters,
we use the empirical SFR calibration for \OII parametrized
in terms of the $B$-band luminosity developed by Moustakas,
Kennicutt \& Tremonti (2006) for local star forming galaxies.
The $B$-band luminosity was regarded as a proxy of stellar mass.
They were aware that the mass-to-light ratios of star forming
galaxies span a wide range and that the $B$-band luminosity
would not be a perfect indicator of stellar mass. However,
they empirically found that the amount of dust extinction
and metallicity are correlated with the $B$-band luminosity
(see Figure 16 in Moustakas et al. (2006)).
They also found that the correlation for star forming galaxies
at 0.5$<$$z$$<$1.5 is roughly consistent with that for
local galaxies. It should be noted, however, that there is a
larger uncertainty in the derived SFR for individual galaxies
at high redshifts.
We apply their correction scheme of dust extinction to our \OII
emitters at $z=1.46$.
We infer the rest-frame $B$-band fluxes from the $J$-band fluxes
and $z-J$ colours and use them to correct for dust extinction.
Figure \ref{fig;sfr_ew}(a) shows thus derived dust-corrected SFRs
of our \OII emitters as a function of distance from the cluster center.

Stellar masses of the \OII emitters are estimated from $K_s$ total
magnitudes using the equation given in \citet{dad04}.
\citet{dad04} estimate the stellar masses for $K$-selected galaxies at
$z>1.4$ using K20 survey data. Mass to $K$-band luminosity ratio is
derived from multi-wavelength data from $U$ to $K$ with \citet{sal55}
IMF (\citealt{fon04}). The relation is calibrated with $z-K$ colour,
which can reduce the dispersion of derived stellar mass to
$\sigma(\Delta \log({\rm M_\star}))=0.20$.
We also derive specific SFR by dividing SFR by stellar mass. Figure
\ref{fig;sfr_ew}(b) shows the specific SFRs for \OII emitters as a
function of radius from a centre.

The observed equivalent width of an \OII emission is also derived from
\OII line flux and continuum flux density
(Equations (\ref{eq;fline}) and (\ref{eq;fcont})).
Figure \ref{fig;sfr_ew}(c) shows the
observed equivalent widths for the \OII emitters as a function of radius
from a centre.

Figure \ref{fig;sfr_ew} suggests that star formation activities of
the \OII emitters in denser regions are as active as those in the
outskirt of the cluster. The equivalent widths are not correlated with
the position of galaxies in the cluster. These facts also suggest
active star formation in the central region of the 2215 cluster. On
the other hand, the specific SFRs may show a mild
correlation in the sense that \OII emitters at the inner regions tend
to have lower specific SFRs. Since the similar correlation is not seen
in their SFRs, this is due to the fact that massive galaxies tend to
reside near the centre of the cluster. 
This tendency may be due to the galaxy formation bias that massive
galaxies are formed in the cluster core, or may be a result of
efficient mass assembly due to merging that can be more effective in
high density regions. 
Figure \ref{fig;sfr_ew}(b) also shows that the red \OII emitters tend
to have lower specific SFRs.
This may be because we underestimate their intrinsic SFRs due to the
strong dust extinction for the red galaxies. Otherwise, this may imply
that these galaxies are just quenching their star formation
activities. 

\subsection{Colour-magnitude diagram}
\label{sec;cmd}

%% figure 9a, 9b
%%%%%%%%%%%%%%%%%%%%%%%%%%%%%%%%%%%%%%%%%%%%%%%%%%%%%%%%%%%%%%%%%
\begin{figure*}
 \includegraphics[width=70mm]{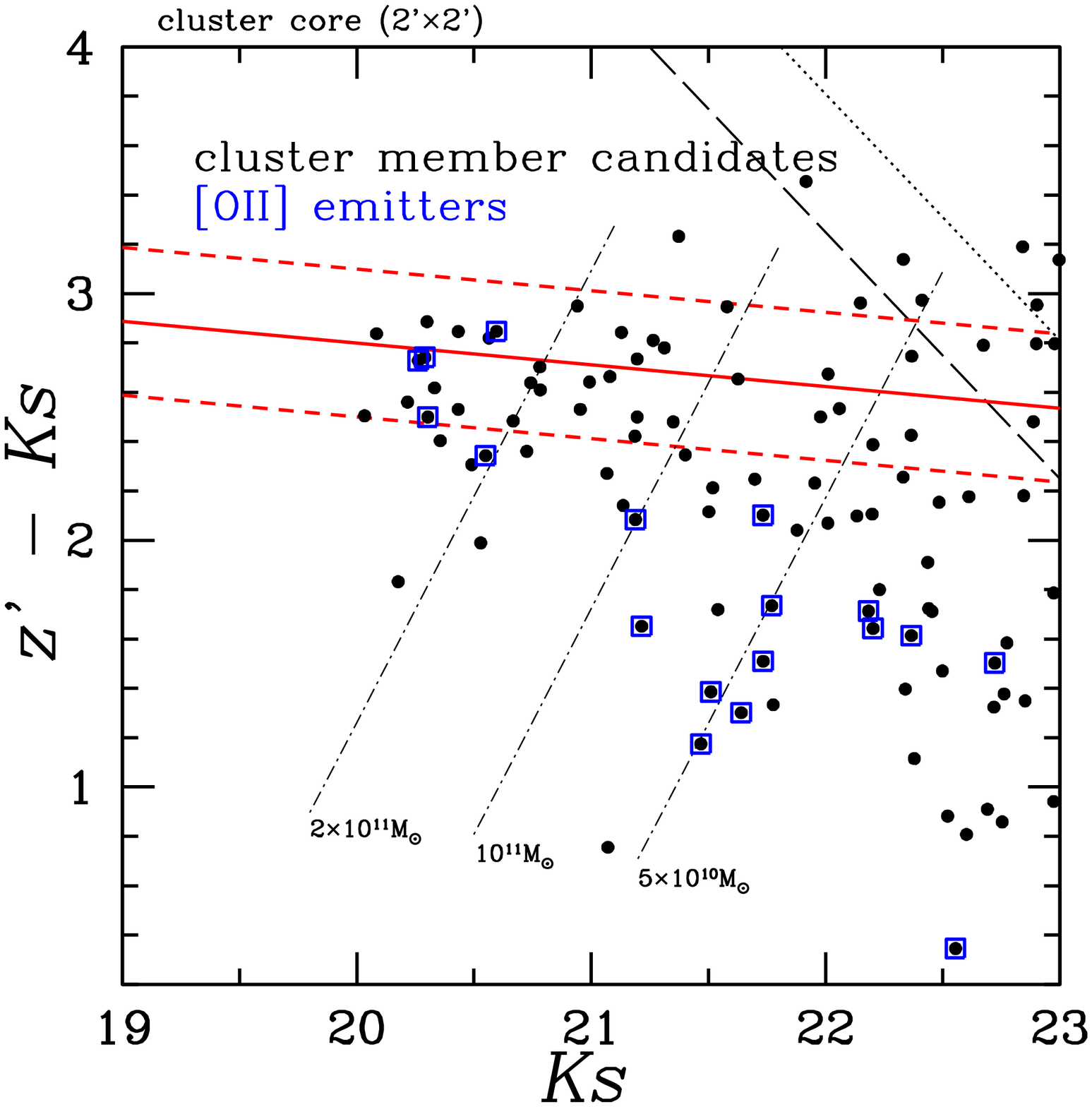}
 \includegraphics[width=70mm]{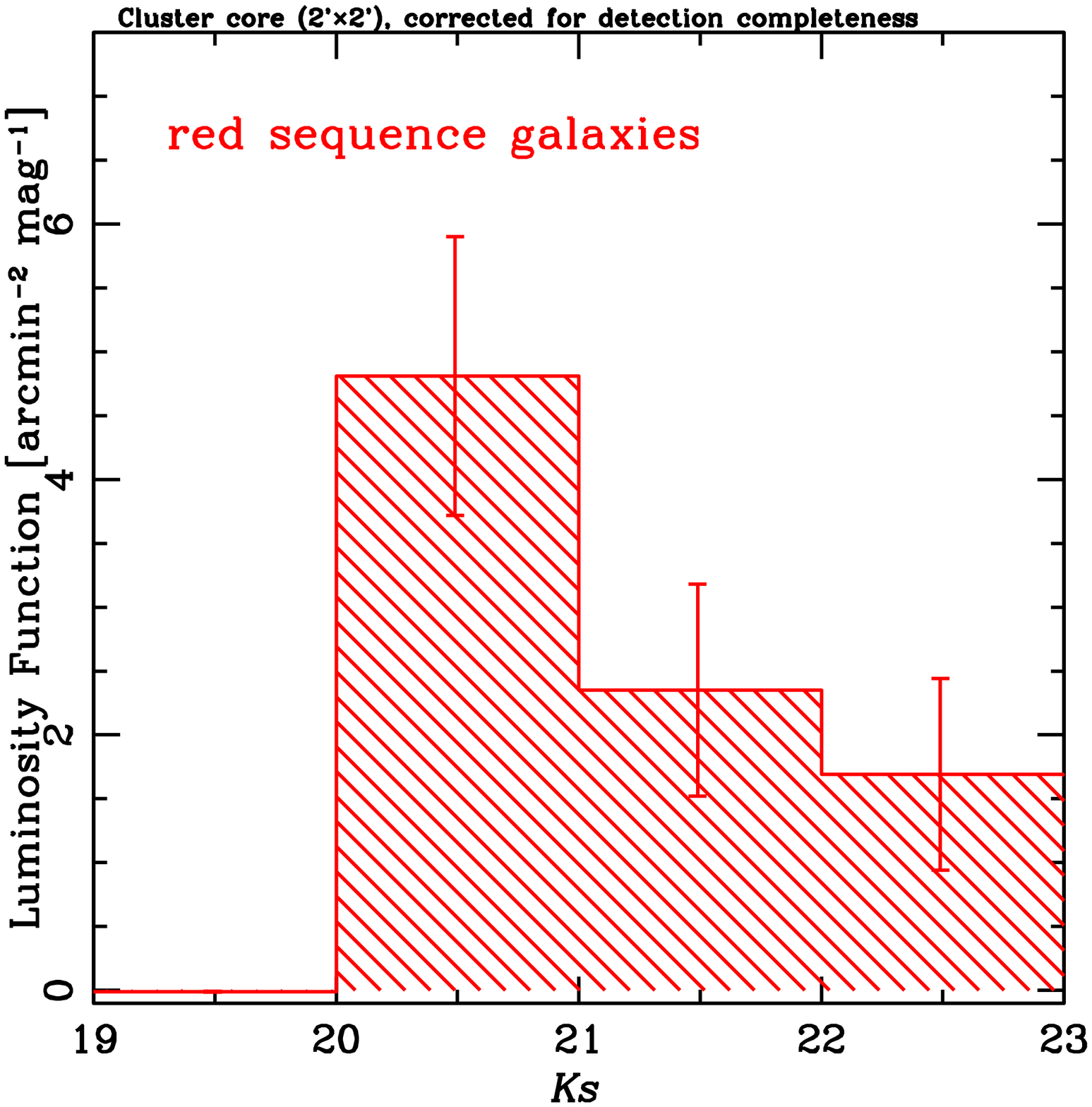}
 \caption{(a) Left panel: Colour-magnitude diagram of $z'-K_s$
   vs. $K_s$ for the $K_s$-detected galaxies in the central
   $2'\times2'$ region.  Black dots shows cluster member candidates.
   Red solid line show the expected location of the colour-magnitude
   relation of passively galaxies galaxies at $z=1.46$ formed at
   $z_{form}$=5 (\citealt{kod98}). 
   We define the red sequence galaxies as those falling between the
   two broken lines, $\Delta(z'-K_s)=\pm0.3$ around the colour-magnitude
   relation.  Long-dashed and dotted lines are 3$\sigma$ and 5$\sigma$
   confidence levels in colours, respectively. 
   Blue open squares show 18 \OII emitters in this region.
   Three dot-dashed lines indicate iso-stellar mass curves for
   $2\times10^{11}$\Msun, $10^{11}$\Msun and $5\times10^{10}$\Msun,
   respectively, which are drawn based on the equation given in
   \citet{dad04}. 
   (b) Right panel: $K_s$-band luminosity function of the red sequence
   galaxies in the central $2'\times2'$ region. 
   The contribution of field galaxies, estimated from the SDF sample,
   is statistically subtracted.  Error-bars show Poisson errors.
   }
 \label{fig;cmd_lf}
\end{figure*}
%%%%%%%%%%%%%%%%%%%%%%%%%%%%%%%%%%%%%%%%%%%%%%%%%%%%%%%%%%%%%%%%%

Figure \ref{fig;cmd_lf}(a) shows a colour-magnitude diagram of $z'-K_s$
vs. $K_s$ for cluster member candidates in the central $2'\times2'$ region.
Note that the $K_s$-selected galaxy sample is used in this section.
The use of the $NB912$-selected sample may cause incompleteness in the
number of galaxies at faint $z'$ magnitudes.
The solid line in the figure shows the expected location of
the colour-magnitude relation (CMR) of passively evolving galaxies
at $z=1.46$ formed at $z_{form}=5$ inferred from the \citet{kod98} model
which is calibrated to reproduce the CMR of elliptical galaxies in the
Coma cluster at $z=0$;
\begin{equation}
z'-K_s=-0.088K_s+4.56.
\label{eq;cmr}
\end{equation}
We define the red sequence galaxies as those falling in-between
$\Delta(z'-K_s)=\pm0.3$ from the predicted CMR as shown by the
broken lines in Figure \ref{fig;cmd_lf}(a).

The number of the red sequence galaxies seems to decrease
at magnitude fainter than $K_s\sim$21.5.
The long-dashed and dotted lines show the 5$\sigma$ and
3$\sigma$ confidence levels of $z'-K_s$ colours, respectively. 
Therefore the decrease in the number of the red galaxies
is not driven by incompleteness. \citet{sta06} and \citet{hil09} also
examine colour-magnitude diagrams, and find that there is a
well defined CMR of red galaxies in the 2215 cluster. 
In \citet{sta06}, however, shallowness of the data does not allow us to
discuss the faint end of the red sequence. \citet{hil09} use deeper data
to investigate the CMR, and colour-magnitude diagram of Figure~6 in
\citet{hil09} shows a deficit of red galaxies at magnitudes fainter
than $K_s\sim21.5$. This is consistent with our result.
 
Since all the cluster member candidates are plotted in Figure
\ref{fig;cmd_lf}(a), some galaxies actually do not belong to the cluster.
In order to correctly evaluate the deficit of the red galaxies,
we derive the field-corrected $K_s$-band luminosity function in the
central $2'\times2'$ region by statistically subtracting the field contamination.
The detection completeness is also corrected for as follows.
Artificial objects with Gaussian profile are randomly
generated and embedded in the raw $K_s$ image, and detection of these objects is
conducted with the same manner as in \S\ref{sec;selection_nbcatalogue}.
The assumed distribution of brightness of the artificial objects is
uniform within each magnitude bin.
The completeness thus estimated in each bin is always as high as $\ga$90\%.

The resulting luminosity function of the red sequence galaxies is shown
in Figure \ref{fig;cmd_lf}(b). It is clear that the number of red member
galaxies decreases at $K_s\la 21.5$.  This magnitude corresponds to
$K_s^*$+0.5 with respect to the passively evolving galaxies at $z=1.46$ \citep{kod98}.
In the 2215 cluster at $z=1.46$, the red sequence is visible
only down to $\sim$$K_s^*$+0.5 and truncated at that magnitude.

Such truncation is seen, if any, at fainter magnitudes ($\ga M^*$+0.5)
in the cores of lower redshift clusters.
For example, \citet{and06} find no deficit of galaxies on red
sequence down to $M^*+3.5$ in the MS1054.4-0321 cluster at $z$=0.83.
\citet{koy07} suggest that the build-up of the CMR depends on X-ray
luminosity of clusters at $z\sim0.8$, and find that the CMR is
established down to $M^*+2.0$ even for a X-ray fainter cluster
RXJ1716+6708 at $z=0.81$.
\citet{lid04} and \citet{tan08} study CMR for the
RDCS~J1252.9-2927 cluster at $z$=1.24.
While the red sequence galaxies only appear down to $K\sim22$
in sub-clumps in the outskirt of the cluster (\citealt{tan08}),
faint red galaxies clearly exist down to $K\sim24$ in the main cluster
(\citealt{lid04}). \citet{lid08} also find that there are red faint
galaxies down to $K\sim24.5$ in the core region of the
XMMU~J2235.3-2557 cluster at $z$=1.39.
We also note that \citet{kod07} show that the CMR of proto-clusters
at much higher redshifts ($2\la z\la 3$) becomes less conspicuous,
and even the bright-end of the red sequence seems to disappear
in proto-clusters at $z\ga 2.5$.

In Figure \ref{fig;cmd_lf}(a), we mark the \OII emitters with blue open
squares.  Most of the \OII emitters have bluer colours as expected since
they are likely to be still forming stars.  The galaxies that are fainter
in $K_s$ hence less massive galaxies tend to be slightly bluer.
This may suggest that star formation activity in more massive galaxies
tends to be truncated at earlier times.
We draw three iso-stellar mass curves of $2\times10^{11}$\Msun,
$10^{11}$\Msun\ and $5\times10^{10}$\Msun, respectively,
using the equation given in \citet{dad04}.
If the \OII emitters cease their star formation, they would
move along these curves until they reach on to the red sequence.
If an \OII emitter has a SFR of 50\Msun/yr, which is close to
the median SFR in our \OII emitter sample, and keeps this rate
constant until $z=1.0$, its stellar mass would increase by
$\Delta$M$_\star$$\simeq7\times10^{10}$\Msun.
If we assume that an \OII emitter gradually becomes red while
increasing its stellar mass at a constant SFR, the faint end of the
red sequence would be filled up by $z\sim1.0$.
If this is the case, the \OII emitters with $\la5\times10^{10}$\Msun\ may
be good progenitors of the faint galaxies on the red sequence.
However, we do not know yet the mechanisms of changing galaxy colours and
quenching their star formation.
As described in the last paragraph of \S\ref{sec;sf_activity},
contribution of AGNs to the \OII emitters on the red sequence can be
large (\citealt{yan06}).  Perhaps AGN feedback is one of the key
mechanisms to reduce the star formation activities.

Combining with the previous studies of CMR (see above), we come up
with the following scenario of formation of red sequence
in clusters at $z\sim1.5$.  The most massive galaxies brighter than
$K_s^*$ (i.e., $>10^{11}$\Msun) are formed at $z>2$, and become red
by quenching the star formation in early epoch.
Some galaxies on the red sequence may still be keeping residual
star formation activities.  On the other hand,
less massive galaxies are actively growing at $z<2$, and we hardly see
galaxies fainter than $K_s^*$ on the red sequence.
Star forming galaxies with $\sim5\times10^{10}$\Msun\ at $z\sim1.5$
may evolve into faint red sequence galaxies by $z\sim1.0$.
This suggest down-sizing propagation of star formation in high redshift
clusters.

\section{SUMMARY}
\label{sec;summary}
We performed a unique, unbiased \OII line survey of star forming galaxies
in the XMMXCS~J2215.9-1738 cluster at $z=1.46$, which is currently the most
distant cluster ever identified with a detection of extended X-ray emission.

We have obtained wide-field optical ($B$, $z'$, $NB912$) and
near-infrared ($J$ and $K_s$) data with Suprime-Cam and MOIRCS,
respectively.
With a combination of $NB912$ narrow-band filter
($\lambda_c$ = 9139\AA, FWHM=134\AA) and the $z'$-band filter,
we detect 69 $NB912$ emitters in the central $6'\times6'$ region
where near-infrared data are also available.
Among them, 44 emitters are
identified as \OII emitters associated to the cluster based on
the $B-z'$ and $z'-K_s$ colours, down to a dust-free star formation
rate of 2.6 \Msun yr$^{-1}$ (3$\sigma$).

We find that many \OII emitters reside in the central high density
region even within a radius of 0.25 Mpc (physical scale).  We also
find that the fraction of \OII emitters to cluster members remains
high up to the core region. This suggests that the 2215 cluster
is still actively forming stars even at the central region, in contrast
to lower redshift clusters, where old passively evolving elliptical
galaxies dominate.
This indicates an inside-out propagation of star formation in high redshift
clusters, and we may be eventually beginning to enter the epoch of biased galaxy
formation in the densest region at $z=1.46$.

SFRs, specific SFRs, and equivalent widths are derived for 44 \OII
emitters.  It is found that the emitters have similar SFRs and
equivalent widths irrespective of the location within the cluster. It
seems however that the specific SFRs tend to decrease slightly toward
the cluster core, probably due to the fact that more massive \OII
emitters exist near the centre.
We may be approaching to the formation phase of massive galaxies
at the cluster core.

Moreover, the colour-magnitude diagram in the 2215 cluster shows a
deficit of red sequence galaxies fainter than $\sim$$M^*+0.5$, while
the red sequence in lower redshift clusters extends to much fainter
magnitudes.
While some bright \OII emitters are located on the red sequence,
all the faint \OII emitters with $>$$M^*$ have blue colours.
It is likely that those blue \OII emitters become redder once they
truncate their star formation, and they would eventually reach and
fill the faint end of the red sequence at lower redshifts. 
This indicates a down-sizing propagation of star formation in high
redshift clusters. 

%%%%%%%%%%%%%%%%%%%%%%%%%%%%%%%%%%%%

\section*{Acknowledgments}
The optical and near-infrared data used in this paper are collected at
the Subaru Telescope, which is operated by the National Astronomical
Observatory of Japan. We thank the Subaru Telescope staff for their
invaluable help to assist our observations with Suprime-Cam and MOIRCS.
We are grateful to Dr. Masayuki Tanaka for carefully reading the
manuscript, and for his valuable comments. We thank Dr. Chun Ly for
kindly providing us with the NB921 emitter catalogue in the Subaru Deep
Field. We also thank Dr. Masafumi Yagi for providing us with some data
reduction codes for the optical imaging data, and for instruction of
the reduction procedures. We would like to thank an anonymous referee
for useful comments and suggestions.
M.H. and Y.K. acknowledges support from the Japan Society for the
Promotion of Science (JSPS) through JSPS Research Fellowship for Young
Scientists. This work was financially supported in part by the
Grant-in-Aid for Scientific Research (Nos. 18684004 and 21340045) by
the Japanese Ministry of Education, Culture, Sports, and Science.

%%%%%%%%%%%%%%%%%%%%%%%%%%%%%%%%%%%%

%%%%%%%%%%%%%%%%%%%%%%%%%%%%%%%%%%%%

\label{lastpage}

\end{document}